\documentclass[lettersize]{IEEEtran}
\usepackage{amsmath,amsfonts}
\usepackage{algorithmic}
\usepackage{array}
\usepackage[caption=false,font=normalsize,labelfont=sf,textfont=sf]{subfig}
\usepackage{textcomp}
\usepackage{stfloats}
\usepackage{url}
\usepackage{verbatim}
\usepackage{graphicx}
\def\BibTeX{{\rm B\kern-.05em{\sc i\kern-.025em b}\kern-.08em
    T\kern-.1667em\lower.7ex\hbox{E}\kern-.125emX}}
\usepackage{balance}

\usepackage[sort&compress,numbers]{natbib}
\usepackage{physics}
\usepackage{multirow}
\usepackage{diagbox}
\usepackage{lipsum}
\usepackage{xcolor}

\begin{document}
\onecolumn
This work has been submitted to the IEEE for possible publication. Copyright may be transferred without notice, after which this version may no longer be accessible.
\twocolumn
\newpage
\title{Evaluation of Machine-generated Biomedical Images via A Tally-based Similarity Measure}

\author{Frank J. Brooks, Rucha Deshpande
\thanks{Manuscript received Month XX, 2025; revised Month XX, 2025; accepted Month XX, 2025. Date of publication Month XX, 2025; date of current version Month XX, 2025. This work was supported in part by the National Institutes of Health (NIH) Awards under Grant P41EB031772.(Corresponding author: Frank J Brooks.)
Frank J Brooks is with Center for Label-Free Imaging and Multiscale Biophotonics, University of Illinois Urbana-Champaign,  Urbana, IL 61801 USA (e-mail: fjb@illinois.edu).
Rucha Deshpande is a research affiliate at the Department of Bioengineering, University of Illinois at Urbana–Champaign, Urbana, IL 61801 USA (e-mail: rmd@illinois.edu).}}


\maketitle

\begin{abstract}
Super-resolution, in-painting, whole-image generation, unpaired style-transfer, and network-constrained image reconstruction each include an aspect of machine-learned image synthesis where the actual ground truth is not known at time of use. It is generally difficult to quantitatively and authoritatively evaluate the quality of synthetic images; however, in mission-critical biomedical scenarios robust evaluation is paramount. In this work, all practical image-to-image comparisons really are relative qualifications, not absolute difference quantifications; and, therefore, meaningful evaluation of generated image quality can be accomplished using the Tversky Index, which is a well-established measure for assessing perceptual similarity. This evaluation procedure is developed and then demonstrated using multiple image data sets, both real and simulated. The main result is that when the subjectivity and intrinsic deficiencies of any feature-encoding choice are put upfront, Tversky's method leads to intuitive results, whereas traditional methods based on summarizing distances in deep feature spaces do not.
\end{abstract}


\section{Introduction}

There is great interest in generative image models (GIMs) throughout the medical imaging community \cite{yi2019generative, koohi2023generative, jung2024image, zhong2024clinical}. The common theme is that an image is partially or wholly synthesized from a model trained on similar image data. Examples include: creating new examples of a rare medical condition as observed via ultrasound imaging \cite{mendez2024leveraging, liang2022sketch} (texture synthesis), digitally transforming low-dose radiography into higher-dose radiography \cite{immonen2022use, kulathilake2023review} (super-resolution), digitally labeling grayscale phase microscopy to look like stained histology images \cite{bai2023deep, rivenson2019phasestain} (unpaired style-transfer), repairing corrupted electron microscopy images \cite{wang2021automatic} (image in-painting), reducing data requirements in image reconstruction via learned constraints \cite{ben2021deep} (learned-prior image reconstruction), and augmenting meager image data sets with new realizations \cite{goceri2023medical, kebaili2023deep, garcea2023data} (whole-image synthesis).

In each use case, the ground truth is not known at the time of application of the pre-trained synthesis model. For example, when a pre-trained GIM is used to in-paint a missing or corrupted region of interest (ROI), the synthesized ROI is \emph{only} plausible. Here, the GIM effectively is doing a kind of high-dimensional imputation, and so the further the unknown truth is from trends observable in the training data, the less accurate the in-painting will be. Another typical use case is when grayscale microscopy images are to be virtually stained histology images via an image-to-image translation network \cite{rivenson2019phasestain, vasiljevic2022cyclegan, bai2023deep}. Note that samples observed in the target histology images often are physically altered by the plating and staining process. Here, the user doesn’t want the phase image to be altered to match the broken or warped structures as in a histology image, but only colored to look similar to some other histology image that would correspond perfectly to the phase image of unadulterated tissue. In both examples, information learned from \emph{other} images is applied to one particular image in order to convey an idea of what's imaged, and not necessarily a genuine attempt at direct pixel-wise mapping between perfectly co-registered images of the same object.

Thus, the problem of evaluating the quality of synthetic images \cite{borji2022pros, theis2016note, betzalel2024evaluation, zhang2018unreasonable} is a problem of measuring the similarity between two images, which are expected to be \emph{only} similar and not the same. Of course, any two arbitrary images might be perceived as similar or dissimilar in numerous ways \cite{mudeng2022prospects, muramatsu2018overview, palubinskas2017image, unnikrishnan2005measures, farhadi2009describing}. In 1977, Amos Tversky published a set-theoretic method for computing the similarity between any two comparands, be they physical, digital, or conceptual \cite{tversky1977features}. The Tversky Index is not a distance in a geometric feature space. Instead, the Tversky Index is based on binary feature exhibition: the user stipulates a feature of interest and then observes whether both comparands exhibit that feature. This is repeated for as many, and as sophisticated of, features as the user likes, then the Tversky Index is computed formulaically to combine all observations into a single number (see Equation~\ref{tversky}). Thus, measuring similarity becomes a tally of common labels. In contrast to distances or divergences, for which ``twice as great'' in no way implies ``half as similar'', a tally has the benefit of always being on a readily interpretable, monotonic, and bounded scale.

Clinical interpretation of images most often is based on the presence of specific radiological signs \cite{ dahnert2011radiology, brant2007fundamentals}, and not on geometric comparison of formulaic feature values \cite{ledley1959reasoning, seising2006vagueness}. For example, a radiologist inspecting a chest radiograph might observe a characteristic ``ground glass texture'' \cite{franquet2001imaging} in the lung and then declare the subject to have pneumonia; this would set the ``pneumonia component'' of a vector possible medical conditions to 1. Other components of such a vector may or may not depend upon each other, but each always is either zero or 1, and the Tversky Index between any two subjects can be readily computed to quantify their a priori defined medical similarity. Here, the radiologist effectively serves as a non-linear transform from an image data space to a medical condition space and at no time are pixel-level image features defined or computed formulaically. That a feature varies substantially across subjects, or might be difficult to describe, is not important. As long as all radiologists can agree on what should count as the feature, then Tversky's method will work.

Tversky's method is well-established and widely applied in a variety of research domains. One apprehension about applying Tversky's method in medical diagnoses is that the subjective choice or assessment of features effectively can change the very definition of a condition which the subjects under comparison might exhibit, or a group to which they may belong. In the context of image comparison, however, classes and conditions are well-defined ab initio. For example, when a human observer is to search for a particular texture, the diagnostic value of that texture is already known. When a machine observer is to perform the same search, an encoding of the texture is already known, as is, too, typically, the fidelity of that encoding. That is, here, the upfront feature specification required by Tversky's method is not a subjective attempt to (re)define a condition or class, but is an evidence-based acknowledgment of what absolutely must be true when the comparands genuinely do share the same condition or class.

It is important to note that Tversky's method does not require features to have any particular meaning or significance. Meaning typically is assigned at the time an image is used. In practice, an observer---human or machine---is assigned an image utility task such as find all lesions in a mammogram, or declare whether a subject has pneumonia from a chest radiograph, and features relevant to that one particular task are specified or learned. For this reason, it is widely understood that any given image might be acceptable to one observer but unacceptable to another. This is directly analogous to how two images can be similar in numerous ways while being dissimilar in numerous other ways. Therefore, it is posited that a feature set useful toward assessing similarity need only be axiomatically relevant to the data and task at hand, and not necessarily the most comprehensive or most generalizable feature set.

At least a few other research groups have attempted to apply versions of the Tversky Index in image analysis problems \cite{salehi2017tversky, abraham2019novel, rahnama2020learning}. In Refs. \cite{salehi2017tversky}  and \cite{abraham2019novel}, the authors attempted to add a kind of Tversky Index to a loss function with the goal of improved image segmentation. The binary feature of interest was a label describing whether a given pixel belonged to the region of interest, and was weighted by the probability of belonging. This is not the sort of use case or implementation we propose. A different group attempted to learn a task-based Tversky Index for comparing images (Ref. \cite{rahnama2020learning}); this is closest to what we propose in that broadly defined image-descriptive features are weighted by relevance to a domain or task. However, those authors effectively inverted the quantification problem by training on paired images that were already deemed similar and then learning the features that made their version of the Tversky Index agree with the known similarity labels.

Alternatively, the present work is focused squarely on demonstrating a systematic procedure for comparing biomedical images where no ground truth is available. The main hypothesis is that Tversky’s method can be used to combine pre-specified, numerically disparate features to provide practically sufficient and robust evaluation of the similarity between any two images, or any two sets of images. The procedure naturally reveals in which features the images compared differ, and thus enables the flagging of individual images for specific deficiencies. This evaluation method is demonstrated for a variety of common scenarios, employing real and algorithmically generated data. For the latter, two novel, visually realistic, feature-tunable stochastic models of fluorescence microscopy are introduced. Finally, caveats on the practical computation and interpretation of similarity measures, including Tversky’s, are discussed.

\section{Proposed Method}

\subsection{Binarization of Feature Values}

The similarity between a subject comparand $\mathcal{S}$ and an archetypical comparand $\mathcal{A}$ is to be computed. Tversky's method requires that features be either present or not. Here, a random variable $X$ is binarized via the indicator function
\[\mathcal{B}(x; l, u) = \begin{cases} 
          		1 & x\in (l, u) \\
          		0 & x\notin (l, u) \\
       			\end{cases}\]
where $(l, u)$ is a tolerance interval which can be derived from the distribution of $X$ as estimated from a set of archetypical images. For example, suppose that every image in an ensemble of synthetic images is to follow the same continuous grayscale intensity distribution as that of an archetypical ensemble. The distribution of the Kolomogorv-Smirnov test statistic can be estimated using a large number of random pairs of archetypes and bounds defined as, for example, the 0.05- and 0.95-quantile values of that test statistic. Of course, numerical representation of object features may require numerous random variables and so each variable necessarily has its own tolerance interval. Thus, a vectorized version of the indicator function, $\mathcal{B}(\vec{X}; \vec{l}, \vec{u})$ yields a vector of the same length as $\vec{X}$, comprising strictly binary components.

\subsection{The Weighted Similarity Index}
The set of all features $F$, regardless of sophistication, exhibited anywhere within two comparands is the union of three mutually exclusive subsets: 1) the set of features found in $\mathcal{S}$ but not in $\mathcal{A}$ ($S\setminus A$), the set of features found in $\mathcal{A}$ but not in $\mathcal{S}$ ($A\setminus S$), and the set of features common to both comparands, ($S\cap A$). Tversky's Index traditionally is defined \cite{tversky1977features}:
\begin{equation}
\textrm{Index}(\alpha, \beta)=\frac{|S\cap A|}{|S\cap A| + \alpha|S\setminus A| + \beta|A\setminus S|}
\label{tversky}
\end{equation}
where $\alpha$ and $\beta$ are user-chosen, non-negative, real scalars describing the relative importance of the set relative complements. Popular choices are $\alpha=\beta=1/2$ and $\alpha=\beta=1$ for which Eq.~\ref{tversky} becomes the Dice Coefficient or Jaccard Index, respectively.

We extend this definition to include a specific weighting of each feature in $F$ by recasting each set as a vector of strictly binary components. Assuming the user has defined $M$ unique features describing the images to be compared, then $\vec{v}_F$ is a length-$M$ vector of ones, $\vec{v}_F=(1,1,\dots,1)$. The subsets $S\cap A$, $S\setminus A$, and $A\setminus S$ are each length $M$ vectors of the form $(f_1, f_2,\dots,f_M)$ where $f_m=\mathcal{B}_m(x_m; l_m, u_m)=0||1$ such that
\begin{equation}
\vec{v}_F = \vec{v}_{S\cap A} + \vec{v}_{S\setminus A} + \vec{v}_{A\setminus S}.
\label{set_sum}
\end{equation}
The Weighted Similarity Index then is
\begin{equation}
\textrm{WSI}(\vec{w})=\frac{\vec{w}\cdot\vec{v}_{S\cap A}}{\vec{w}\cdot\vec{v}_{S\cap A} + \vec{w}\cdot\vec{v}_{S\setminus A} + \vec{w}\cdot\vec{v}_{A\setminus S}}
\label{weighted_tversky}
\end{equation}
where $\vec{w}$ is a length $M$ vector of user-defined weights describing the relative importance of features to the definition of similarity in the particular data domain and task context. Note that each $w_m\in\mathbb{R}\ge1$. This is because if $w_m\in(0,1)$, one could simply renormalize all $w_m$ to the minimum weight and if $w_m = 0$, then that feature should never have been included in $F$ (i.e., there's no reason include features known to be unimportant to the observer). Under this definition, $\textrm{WSI}(\vec{w})$ is a real scalar in $[0,1]$, as in Equation~\ref{tversky}, with $\textrm{WSI}(\vec{w})=0$ indicating ``as practically dissimilar as can be'' and $\textrm{WSI}(\vec{w})=1$ indicating ``as practically similar as can be.'' Numerical features and the feature-weighting vector are defined at time of need.  

\section{Studies, Data, and Methods}

\begin{table}[ht]
\begin{center}
\caption{A summary of studies and data.}
\label{tab:summary}
\begin{tabular}{p{2cm} c p{2cm} c} \hline
    Domain & Data & Study & Comparands \\ \hline 
	Fluorescence \par microscopy & CoBaLT & Whole-feature ablation & Ensembles \\
	Fluorescence \par microscopy & WonoST & Feature \par perturbation & Ensembles \\
	Virtual \par mammography & VICTRE & Quality of\par generated images & Pairs \\
	Chest \par radiography & CheXpert & Reconstruction\par quality & Pairs/Ensembles \\ \hline 
\end{tabular}
\end{center}
\end{table}

\subsection{Stochastic Image Models of Fluorescence Microscopy}

\begin{figure*}[bth]
\begin{center}
\includegraphics[width=0.95\textwidth]{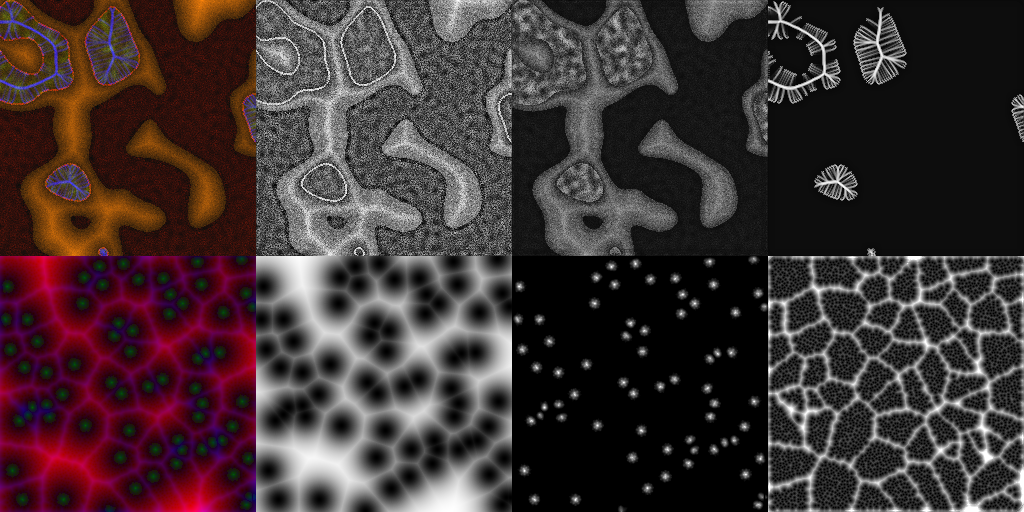}
\caption{Realizations from the stochastic image models of fluorescence microscopy. Top: the correlated-background lumpy triple (CoBaLT) model Bottom: Worley-noise soft tissue (WonoST) model. The RGB composite is shown in the first column and each subsequent column is the view in only the red, green, or, blue channel. The contrast of each channel has been enhanced for display.}
\label{fig:sim_examples}
\end{center}
\end{figure*}

An example from the correlated-background lumpy triple (CoBaLT) model, is shown in the top row of Figure~\ref{fig:sim_examples}; numerous example images are made publicly available. In brief, a realization from a standard lumpy background \cite{rolland1992effect} is thresholded to create background, foreground, and ``cell'' masks. Texture is added to each region by difference of Gaussians applied to random noise. Structure is added to the foreground region (yellow-orange) by computing the Euclidean distance map (EDM). The spiny structure within cells is computed from the largest eigenvalues of the Hessian matrix of the EDM of the cell masks. Edge masks (e.g., the bright coral color) are made from differences of mask and dilation and erosion. The masked regions are combined purposely into a single RGB image such that channel selection perfectly segments every instance of key visual features (e.g., the ``spines'' are exclusively blue); this is convenient for studying the impact of whole-feature ablation on the weighted similarity index (Section~\ref{sec:ablation}). The CoBaLT model runs on the order of seconds on a modern desktop computer, is visually realistic of the \emph{kinds} of features seen in fluorescence microscopy are, is highly tunable, and can generate images of any dimensions. For most experiments, 256x256 was chosen for convenient analysis and data storage.

An example from the Worley-noise soft tissue (WonoST) model is shown in the bottom row of Figure~\ref{fig:sim_examples}. As before, numerous example images are available online REF. In brief, the red channel is the first-nearest-neighbor component of Worley noise \cite{worley1996cellular} computed for a set of Poisson-distributed seeds. The blue channel is distinct first-nearest-neighbor Worley noise pattern computed from a greater set comprising the original seeds and many more. The green channel is a thresholding of the red channel designed to highlight the location of the original seeds while preserving the blue-channel pattern. The channels together exhibit the \emph{kinds} of features seen in soft tissue---compare, for example, the membrane-like structure in the red channel to the images \cite{idrdata} supporting Ref. \cite{hartmann2020image}---but no attempt was made to model any particular tissue. For all experiments, 256x256 was again chosen for convenience.

One experiment requires that visually obvious features be perturbed systematically in image space. The red channel was perturbed by first normalizing the intensity and then histogram-transforming it to a Beta($\alpha$=2, $\beta$=4.4) distribution, which was observed to be the average distribution across the image ensemble. The variance and $\beta$ were held fixed and $\alpha$ computed such that the median intensity would increase by the prescribed percentage. The intensity was then transformed to these new distribution, and mapped back to an 8-bit gray scale. Seed-like structures were removed from the green channel as follows. The green channel was binarized at a threshold of zero. All seed objects enumerated, and a percentage of them removed by setting all pixels with randomly selected seed labels to zero. The membrane structure of the blue channel was perturbed similarly. Seed locations near the centers of the smallest cells were identified via blurring and binarization at the Otsu threshold. For this particular model, the result is a set of single-pixel points for which the Worely noise is computed. A percentage of these points were randomly omitted with probability scaled to the size of the \emph{large-scale} cell, i.e., the cell comprising the smaller ones; this was done to prevent decimation of tiny regions. The first-nearest-neighbor distance between the remaining seeds was recomputed. The result is the same texture as before but with some of the smallest regions appearing a bit larger (see Figure~\ref{fig:worley_examples} in the results).

\subsection{Comparison of Real Medical Images}

Chest radiographs from the CheXpert training dataset \cite{irvin2019chexpert} were selected and processed as follows. CheX\underline{B}ert labels \cite{smit2020combining} were used to select only candidate patients not having any medical devices and labeled either ``pneumonia'' or ``no finding.'' When multiple studies of the same patient were available, only the first study was retained. When multiple views were available, only the last frontal view was retained. Both anterior-to-posterior posterior-to-anterior imaging were retained equally. Because the size of the radiographs varies greatly, only the most common size among our cohort, 2320x2828 pixels, were retained. The remaining patients were balanced via random selection to have equal numbers of males and females. The final cohort comprises $N_S$=1144 cases of unambiguous findings of pneumonia and $N_A$=3724 examples of no finding.

Plausibly corrupted versions of real image data (CheXpert) were created using a variant of two-dimensional principal component analysis (2DPCA) \cite{zhang20052d, nhat2005two}. A two-dimensional technique was chosen because the real images  are too large for one-dimensional analysis, which necessarily includes a flattening operation and then inversion of a large covariance matrix. A method similar to those reported in Refs. \cite{zhang20052d} and \cite{nhat2005two} was implemented as follows. Each image $I_n\in\mathbb{R}^{R\times C}$ is first thought of as a data matrix comprising $R$ realizations of length-$C$ random variable. The covariance matrix of these data is
\begin{equation}
\Sigma_\mathcal{R} = \frac{1}{R}(I_{n}-\langle\mu_n\rangle_r)^{\top}(I_{n}-\langle\mu_n\rangle_r)
\end{equation}
where $\langle\mu_n\rangle_r$ is an $R\times C$ matrix where each row is identically the single vector mean computed along the columns of $I_n$. That is, the mean of the random variable is repeated $R$ times. The average covariance matrix over all images then is:
\begin{equation}
\overline{\Sigma}_\mathcal{R} = \frac{1}{N}\sum_{n=1}^{N}(\Sigma_\mathcal{R})_n
\label{cov_rows}
\end{equation}
Analogously, the same image also may be thought of as a data matrix comprising $C$ realizations of length-$R$ random variable, for which a distinct covariance matrix is computed, that also can be averaged over the ensemble of images. Thus,
\begin{equation}
\overline{\Sigma}_\mathcal{C} = \frac{1}{NC}\sum_{n=1}^{N}(I_{n}-\langle\mu_n\rangle_c)^{\top}(I_{n}-\langle\mu_n\rangle_c)
\label{cov_cols}
\end{equation}
where $\langle\mu_n\rangle_c$ is an $R\times C$ matrix where each column is identically the single vector mean computed along the rows of $I_n$. For clarity, it is noted that $\overline{\Sigma}_\mathcal{R}$ is a $C\times C$ matrix and $\overline{\Sigma}_\mathcal{C}$ is an $R\times R$ matrix and that our average covariances are not the same as those given in Ref. \cite{zhang20052d}. However, as in Ref. \cite{zhang20052d}, we compute projection matrices $P_\mathcal{R}$ and $P_\mathcal{C}$ as matrices comprising the eigenvalue-ordered eigenvectors of $\overline{\Sigma}_\mathcal{R}$ and $\overline{\Sigma}_\mathcal{C}$, respectively. The matrix comprising \emph{vector} loadings can be computed as $L_n = P_\mathcal{C}^{\top}I_n P_\mathcal{R}$. Thus, $L_n$ is an $R\times C$ matrix of expansion coefficients. An image can be reconstructed using any subset of the eigenvectors by including binary matrices $B_\mathcal{R}$ and $B_\mathcal{C}$ to effectively omit entire rows and/or columns of $L_n$. Thus, the reconstructed image is 
\begin{equation}
\widehat{I}_n = P_\mathcal{C}B_\mathcal{C} L_n B_\mathcal{R}^{\top}P_\mathcal{R}^{\top}.
\label{eq:recon}
\end{equation}

One can design a variety of interesting feature-representation errors, directional streaking, and other aliasing artifacts via thoughtful choice of $B_\mathcal{R}$ and $B_\mathcal{C}$. Note that the reconstructed image must be rescaled to the original grayscale to compensate for the contribution (positive or negative) of omitted components. The original CheXpert data were considered to be the ground truth objects from which $\overline{\Sigma}_\mathcal{R}$ and $\overline{\Sigma}_\mathcal{C}$ were computed. High-dose x-ray images were simulated by choosing 1024 components at random with probability inversely proportional to the rank of the eigenvector such that some high frequency components are selected and some lower frequency components are omitted. Mid- and low-dose versions of the same image were created by choosing only 512 or 256 non-zero components, respectively, of $B_\mathcal{R}$ and $B_\mathcal{C}$. The result is three sets of real medical image data that can be compared as if they were imaged via different systems, or output by different generative models. We note that these data are schematic only and that calibration of the corruption to a specific radiation dosage, imaging system, or generative model is neither necessary nor intended.

\subsection{Analysis of Grayscale Images}

In several experiments, we compare the proposed similarity metric to traditional measures. Where appropriate, grayscale images were analyzed by first segmenting them into background, foreground, and ``special'' feature region within the foreground. Note that, for some images, it is possible for any of these ROIs to exhibit holes or comprise disconnected sub-regions. The morphology of each greater ROI is represented as a single: percentage of image size, perimeter-to-area ratio, convexity, and solidity.  The image texture within each greater ROI was represented as asymmetric, normalized gray-level cooccurrence matrices (GLCM) \cite{haralick1973textural} computed at four lengths, which are specified at time of need, and through each orientation $\theta=0^\circ,45^\circ,90^\circ$ and $135^\circ$. Grayscale values were first equal-probability quantized to $K$=16 levels. Known GLCMs were computed from a distinct set ensemble of archetypical images and the texture summarized as the intra-level correlation (formula 3 in the Appendix of Ref. \cite{haralick1973textural}). The analysis procedure yields a set of 4 morphology + 4(4 co-occurrence) = 20 random variables for assessing each of the 3 ROIs such that each image is represented as a point in a 60-dimensional space of interpretable features.

\subsection{Traditional Geometric Comparison}

Embedding images into geometric feature spaces was accomplished using the open-source Img2Vec library \cite{img2vec}. Img2Vec applies any one of a number of pre-trained models to extract a feature vector from an input RGB image. For convenience, we chose the maximally compressed default: a ResNet-18 architecture \cite{he2016deep} pre-trained on the ImageNet database \cite{deng2009imagenet}. Each image in each ensemble analyzed was encoded as a point into a 512-dimensional feature space and the parameters of a multivariate Gaussian model fit to those points. In other words, each \emph{embedded} image is treated as a length-512 variate drawn from a multivariate Gaussian distribution with mean vector $\vec{\mu}$ and covariance matrix $\Sigma$ determined empirically from the feature-space point cloud. From these embeddings, Kullback-Leibler divergence, Mahalanobis distance, or Euclidean distance can be computed, as needed.

\subsection{Computations}

All computations were done in various human-written Jupyter notebooks via Python 3.11.7 running on an iMac M3 computer with 24 GB of RAM running MacOS v14.5. Other libraries employed are: NumPy v1.26.1, SciPy v1.10.1, PIL v9.5.0, scikit-image v0.24.0, Seaborn v0.12.2, and Pandas v1.5.3.

\section{Results}

\subsection{Whole-Feature Ablation}\label{sec:ablation}

\begin{figure}[bth]
\begin{center}
\includegraphics[width=0.475\textwidth]{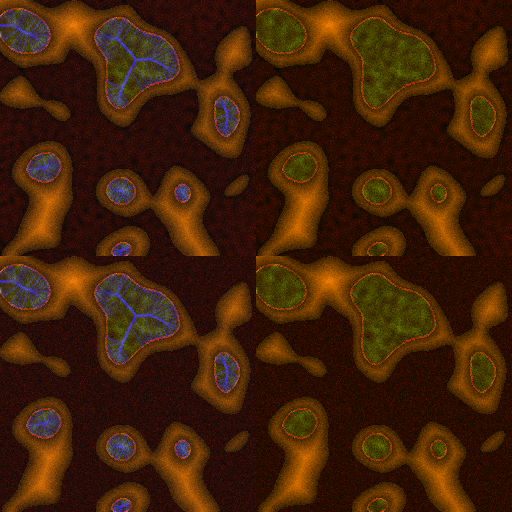}
\caption{Examples of whole-feature ablation in the CoBaLT stochastic context model. Top left: an unadulterated realization. Top right: same realization with the spiny structures removed. Bottom left: same realization with a scrambled background texture. Bottom right: same realization with both spiny structures removed and background texture scrambled.}
\label{fig:cobalt_examples}
\end{center}
\end{figure}

Perhaps the simplest demonstration of how geometric measures of similarity can be inefficacious is to completely remove sophisticated, visually stark features. Realizations of the unadulterated CoBaLT model are defined to be the archetypes and various subjects are defined as follows. The first subject ensemble is defined to be distinct realizations of the CoBaLT model but with the blue channel set to zero---this exactly removes only the spiny structures within the smaller, cell-like ROIs. A second subject ensemble is defined by shuffling the pixels in only the background of the red and green channels---this removes only the background texture. A third subject ensemble is defined by performing both background shuffling and removal of the blue spines. An example from each distribution is shown in Figure~\ref{fig:cobalt_examples}. The universal feature set was defined to be $F$=$\{$correct background texture observable, correct spines observable, all other features exhibited$\}$ and the weighting vector set to $\vec{w}$=(1,1,1). The weighted similarity index (WSI) computed via Equation~\ref{weighted_tversky} is shown in Table~\ref{tab:wsi_for_cobalt_ablation} for the three distinct subject ensembles compared to the archetypes.

\begin{table}[ht]
\begin{center}
\caption{The weighted similarity index computed via Equation~\ref{weighted_tversky}.}
\label{tab:wsi_for_cobalt_ablation}
\begin{tabular}{c c c c}
    ~ & missing texture & missing spines & missing both \\
    $\vec{v}_{S\cap A}$ & (0,1,1) & (1,0,1)& (0,0,1) \\
    $\vec{v}_{S\setminus A}$ & (1,0,0) & (0,1,0) & (1,1,0) \\
    $\vec{v}_{A\setminus S}$ & (0,0,0) & (0,0,0) & (0,0,0) \\ \hline
    WSI & $2/3$ & $2/3$ & $1/3$ \\ 
\end{tabular}
\end{center}
\end{table}

The similarity between the same archetype and subject ensembles was assessed by first embedding each image into a 512-dimensional deep feature space and then computing the Kullback-Leibler divergence as described in the methods. The results are shown in Table~\ref{tab:kld_feature_ablation} where there are several key observations to be made. First, the percent inter-quartile range in the \emph{intra}-distribution KLD computed across folds---these are the uncertainties along the diagonal in Table~\ref{tab:kld_feature_ablation}---increases as whole features are removed. This makes sense for this stochastic image model because the non-random features constrain the total randomness. That is, for example, removing the spatial correlation of the background adds variance to each realization. Second, the median KLD within each distribution is about the same, which indicates that the encoding scheme is effective. Third, some divergences between unequivocally different ensembles---e.g., the unadulterated version and the version with the background texture missing---are \emph{less} than the intra-ensemble divergence. This is a clear example of the distance-based method failing to distinguish mundane variance from meaningful variance. Fourth, note that the sizes of divergence differences are not intuitive. For example, when using the ``missing spines'' distribution as the archetype (third column), the unadulterated distribution is about twice as great from the intra-ensemble divergence but the ``missing texture'' distribution is about 3 times as great. In rank these factors make sense but, given the relative size and complexity of the missing spines to those of the missing texture, it is not obvious why one divergence is 50\% greater than the other, or that a 50\% increase could be used to infer anything about the images. That is, the geometric measure seems to ambiguously reflect visually obvious differences. 

\begin{table}[ht]
\addtolength{\tabcolsep}{-4pt}
\begin{center}
\caption{Kullback-Leibler divergence between the subject (row) and archetype (column) point distributions.}
\label{tab:kld_feature_ablation}
\begin{tabular}{|c| c c c c|}\hline
    missing: & nothing & texture & spines & both \\ \hline
    nothing & 276.9 $\pm$ 0.831\% & 106.9 & 542.7 & 579.3 \\
    texture & 109.4 & 277.4 $\pm$ 1.39\% & 903.8 & 545.8 \\
    spines & 250.1 & 420.2 & 277.0 $\pm$ 2.1\% & 116.6 \\
    both & 359.8  & 280.0 & 128.2 & 278.4 $\pm$ 3.04\% \\ \hline
\end{tabular}
\end{center}
\end{table}

\subsection{Systematic Feature Perturbation}

Recognizable features in $N$=2048 realizations of WonoST model with exactly 64 Poisson-distributed seeds were systematically perturbed as described in the Methods. The level of perturbation is computed as a percentage increase. At each level, the KLD is computed between the perturbed distribution (the subject) and an unperturbed distribution of equal size (the archetype). A paired archetype is the exact same images which were perturbed; and unpaired archetype is a new set unperturbed images not seen during perturbation. Examples of perturbed images are shown in Figure~\ref{fig:worley_examples}. The divergence results are shown in Table~\ref{tab:kld_feature_reduction}.

\begin{figure}[bth]
\begin{center}
\includegraphics[width=0.475\textwidth]{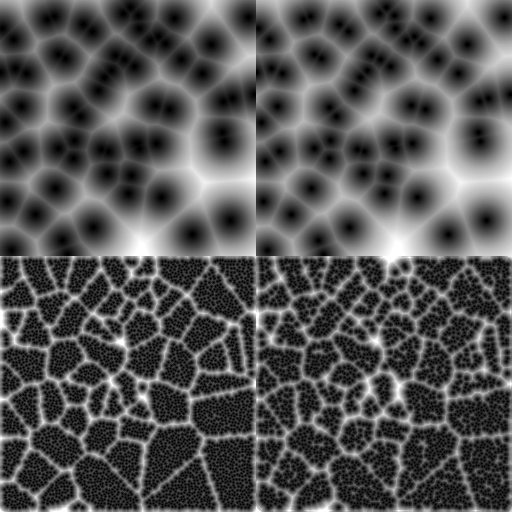}
\caption{Examples of feature reduction in the WonoST stochastic context model. Top: red channel. Bottom: blue channel. Perturbation of the green channel is straightforward and needs no illustration. The left column is the unperturbed channel and the right column is the largest perturbation (32\%). Note the many, subtle differences observable via close inspection. All images have been contrast enhanced for display.
}
\label{fig:worley_examples}
\end{center}
\end{figure}

The red channel, which corresponds to the stark, long-range membrane-like structure, was perturbed by progressively increasing the median intensity while holding the variance fixed. It is seen that even a small change (2-4\%) in intensity is reflected by a non-zero KLD, indicating that the deep features and multivariate Gaussian model of them are sufficiently sensitive for our demonstrations. This is seen for all single-channel perturbations. The green channel was perturbed simply by removing $p\%$ of the seeds at random. The blue channel was perturbed similarly by removing $p\%$ of the smaller-scale cell centers at random at recomputing the first-nearest neighbor Worley noise (which is how the blue channel is computed; see Methods). The scaling of the KLD is approximately quadratic for each channel, however, the actual rate of change differs across channel.

Typically, paired ground-truth data are not available so perturbed image ensembles were also compared to new set of unperturbed data from the same stochastic model. The same quadratic scaling of divergence is seen for the comparisons between unpaired ensembles, however, the values are much greater than for the paired data (Table~\ref{tab:kld_feature_reduction}). This indicates the intra-ensemble variance among like images is much greater than the inter-ensemble variance caused by perturbation. The upshot is that even when statistically robust, visually unmissable changes are made to every image in the subject ensemble, the KLD divergence barely increased.

This is in stark contrast to the WSI for the same data. Here, the median intensity, the number of seeds, or the number of distinct cells \emph{are} the features. One way to consider all types of perturbation at once is to set $\mathcal{F}=$\{$f_R$=red intensity correct, $f_G$=green number of seeds correct, $f_B$=blue number of cells correct\} and $\vec{w}=(1,1,1)$. A threshold that defines acceptability can be chosen for each channel, for example, $\mathcal{B}({\mathcal{F})}=(f_R\pm10\%, f_G\pm<5\%, f_B\pm<20\%)$. Under these stipulations, the weighted similarity index is given in the last column of Table~\ref{tab:kld_feature_reduction}.

\begin{table}[ht]
\begin{center}
\caption{Kullback-Leibler divergence between perturbed (subject) and unperturbed (archetype) distributions.}
\label{tab:kld_feature_reduction}
\begin{tabular}{|c|c c c | c c c | c |} \hline
	\multirow{2}{*}{\%}  & \multicolumn{3}{c}{Paired} & \multicolumn{3}{c}{Unpaired} & WSI\\ \cline{2-8}
    ~ & red & green & blue & red & green & blue & ~ \\ \hline
    2  & 0.4774 & 4.920 & 3.813 & 103.1 & 102.4 & 101.5 & 1.0000 \\
    4  & 1.0240 & 16.10 & 8.780 & 103.7 & 106.7 & 103.5 & 1.0000 \\
    8  & 3.4810 & 31.74 & 21.47 & 104.9 & 114.1 & 109.1 & 0.6667 \\
    12 & 6.9200 & 62.88 & 38.13 & 107.5 & 133.2 & 116.7 & 0.3333 \\
    16 & 11.630 & 87.27 & 57.01 & 111.9 & 150.6 & 128.4 & 0.3333 \\
    24 & 23.970 & 185.7 & 103.4 & 120.6 & 226.4 & 162.9 & 0 \\
    32 & 40.140 & 348.2 & 166.5 & 133.6 & 367.2 & 214.0 & 0 \\ \hline
\end{tabular}
\end{center}
\end{table}

\subsection{Assessing Generated Image Quality} 

Individual image comparison is demonstrated using three distinct sets of  images ($N$=$10^4$ in each set) generated using models submitted as part of a recently reported AAPM Grand Challenge \cite{deshpande2025report}. The challenge was to generate accurate realizations of random cross-sections of the VICTRE breast phantom \cite{badano2018evaluation}. The sets were chosen subjectively by the current authors to be so visually different that the quality ranking is axiomatic. \emph{It is explicitly noted:} what follows is only a demonstration and no claim is made that the generated images or analyses thereof are fit for any particular purpose.

A random forest classifier (training set $N$=5000, 64 trees, with bootstrapping, and no restrictions on branching) was trained to use 60 conventional feature values (see Methods) to predict the human-observed ranking (test set $N$=5000, observed accuracy $>99$\%). The top 16 most important variables (i.e., features) were retained for further analysis. The component of the feature-weighting vector corresponding to the least important variable was set to 1 and all other components set proportionally to variable importance. Each feature tolerance was set to $\pm20\%$ of the archetype feature value.

\begin{figure}[bth]
\begin{center}
\includegraphics[width=0.475\textwidth]{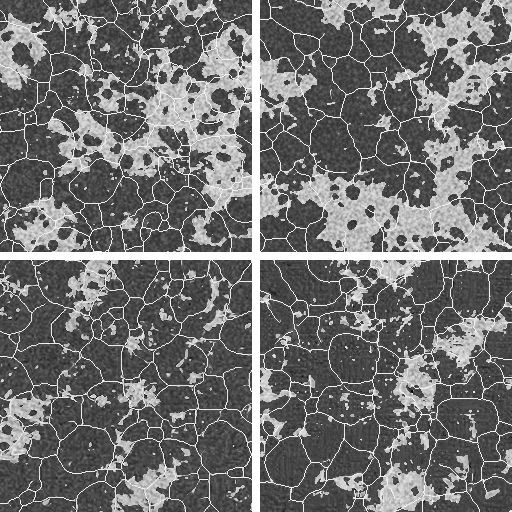}
\caption{Examples of similar images submitted to the AAPM Grand Challenge on deep generative models \cite{deshpande2025report}. Each image is a 256x256-pixel region taken from a larger elliptical shape. At the top-left is training data, top-right is Rank 1, bottom-left is Rank 2 and bottom-right is Rank 3; these ranks are subject to the present work and do not correspond with the contest ranking. Differences from the training image in the bright ``skeleton,'' the foreground shape, and texture are obvious throughout the ranks.
}
\label{fig:worley_examples}
\end{center}
\end{figure}

\subsubsection{Ensemble Cross-similarity}

\begin{figure}[bth]
\begin{center}
\includegraphics[width=0.45\textwidth]{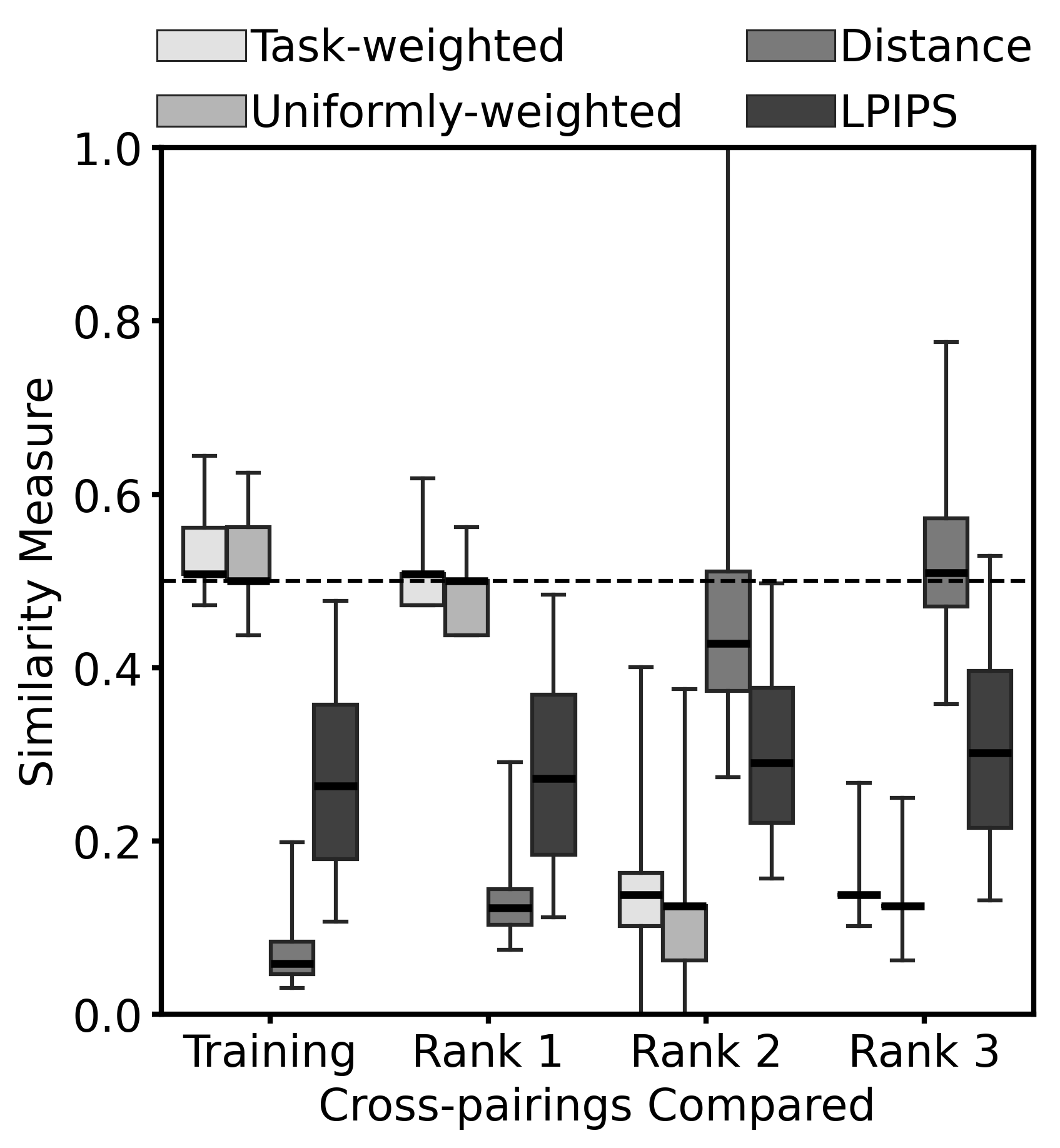}
\caption{Boxplot of the similarity between random pairs comprising a subject image (x-axis) and a training image. The Rank 1 has nearly the same range of values as does comparing training data to other training data (first boxes). Rank 2 has much lower similarity, on average, implying that the ensemble comprises many images with features not seen in the training data. Essentially none of the similarity-defining features are seen in Rank 3 images. The clear distinction between the ensembles is not obvious from the distance alone.}
\label{boxplot_inter-wsi}
\end{center}
\end{figure}

The archetype image ensemble was declared to be a size $N$=10000 random sample of the training data published for the AAPM Grand Challenge \cite{gotsis2023data}. The subject ensemble was set to be each of the three human-ranked sets in turn. An archetype-subject pair was selected at random and both the task-weighted and uniformly weighted similarity index computed. For comparison, the Euclidean distance between the feature space points, and the learned perceptual image patch similarity (LPIPS) \cite{zhang2018unreasonable} between whole images, were computed as well. This procedure was repeated for $N$=32768 unique pairs, for each ranked set. A boxplot of the results is shown in Figure~\ref{boxplot_inter-wsi}. Foremost, weighting clearly impacts the value of similarity; but we note that, in our example, the results are qualitatively the same regardless of weighting and that the choice, impact, and interpretation of weighting is always specific to the application. It is also clear from Figure~\ref{boxplot_inter-wsi} that the submitted ensembles are very different. According to the WSI, the human-ranked-1 entry is nearly as similar to the training data as the training data are to themselves. The Rank-2 entry is about five times less similar to the training data. The Rank-3 entry is clearly the poorest performer with almost all similarity values indicating that only one feature is shared with the training data. In contrast, the scale of the distance is not obviously interpretable. For example, Rank-1 is about twice the expected intra-training distance, despite both visual inspection and the similarity index indicating roughly equal quality. Rank-2 is about eight times the value and Rank-3 is about 9 times the value. So, by distance, Rank-1 seems worse that it is while ranks 2 and 3 seem of comparable quality, which, visually, they certainly are not. The LPIPS has the correct sign of effect in that it increases as visual quality decreases, however, that increase is slight and the LPIPS values are approximately the same range for all ranks. Thus, the WSI agrees well with visual intuition whereas the distance-based measures do not.

\subsubsection{Ensemble Self-similarity}
\begin{figure}[bth]
\begin{center}
\includegraphics[width=0.45\textwidth]{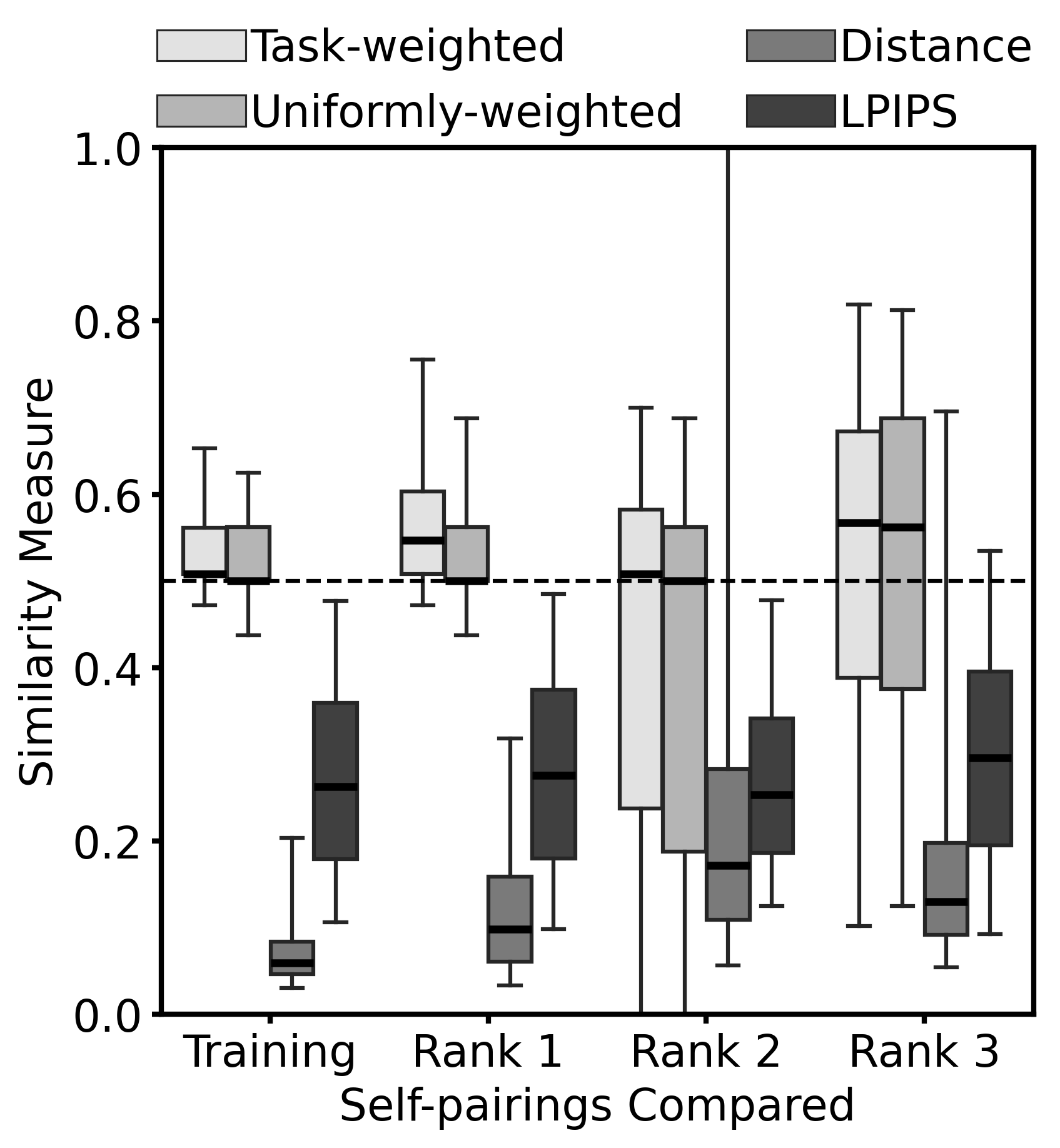}
\caption{Boxplot of the similarity between random pairs of images from the same ensemble. Rank 1 exhibits the same median value of self-similarity as do the training data. Rank 2 exhibits many images that are atypically dissimilar, which indicates that the kind of images that ought to be made are not being made. Rank 3 exhibits many images that are atypically similar, which indicates that some types of images are being made too often, which is a kind of memorization. None of these nuances are obvious from the distance alone.}
\label{boxplot_intra-wsi}
\end{center}
\end{figure}

The procedure from the previous example was repeated but this time with the subject and archetype chosen at random from the same ensemble. The results are shown as a boxplot in Figure~\ref{boxplot_intra-wsi}. The dashed line indicates the median value of the uniformly weighted similarity index and the gray band indicates the outlier range ($<2.5$\%, $>97.5$\%). Values above the band indicate that realizations are more often similar to each other than is expected from the training data; this an indicator of memorization. Values below the band are less similar than expected from the training data; this indicates the DGM is not creating the \emph{kind} of images that it should. That is, Rank-2 and Rank-3 both include images that are different from any seen in the training data. Furthermore, the length of the boxes---i.e., the interquartile range---indicates that Rank-2 entry more often includes these images than does the Rank-3 entry. Incidentally, this observation is consistent with the published report \cite{deshpande2025report} as our Rank-2 entry included many generated images with a kind of ``high-frequency feature sticking.'' Because our stipulated Rank-2 ensemble is less diverse in that it makes essentially the same prominent error often, it more often tests as dissimilar in comparison to some of the more diverse ensembles. Additionally, the median value for the Rank-3 entry is clearly greater than expected for the training data. This indicates that the Rank-3 ensemble exhibits less variation overall in that it is too often, too strongly similar to itself. None of these observations are obvious from either of the distance-based measures alone.

\subsection{Analysis of Real Medical Images}

\subsubsection{Comparing Ensembles of Different Diagnoses}

\begin{figure*}[bth]
\centering
\includegraphics[width=0.95\textwidth]{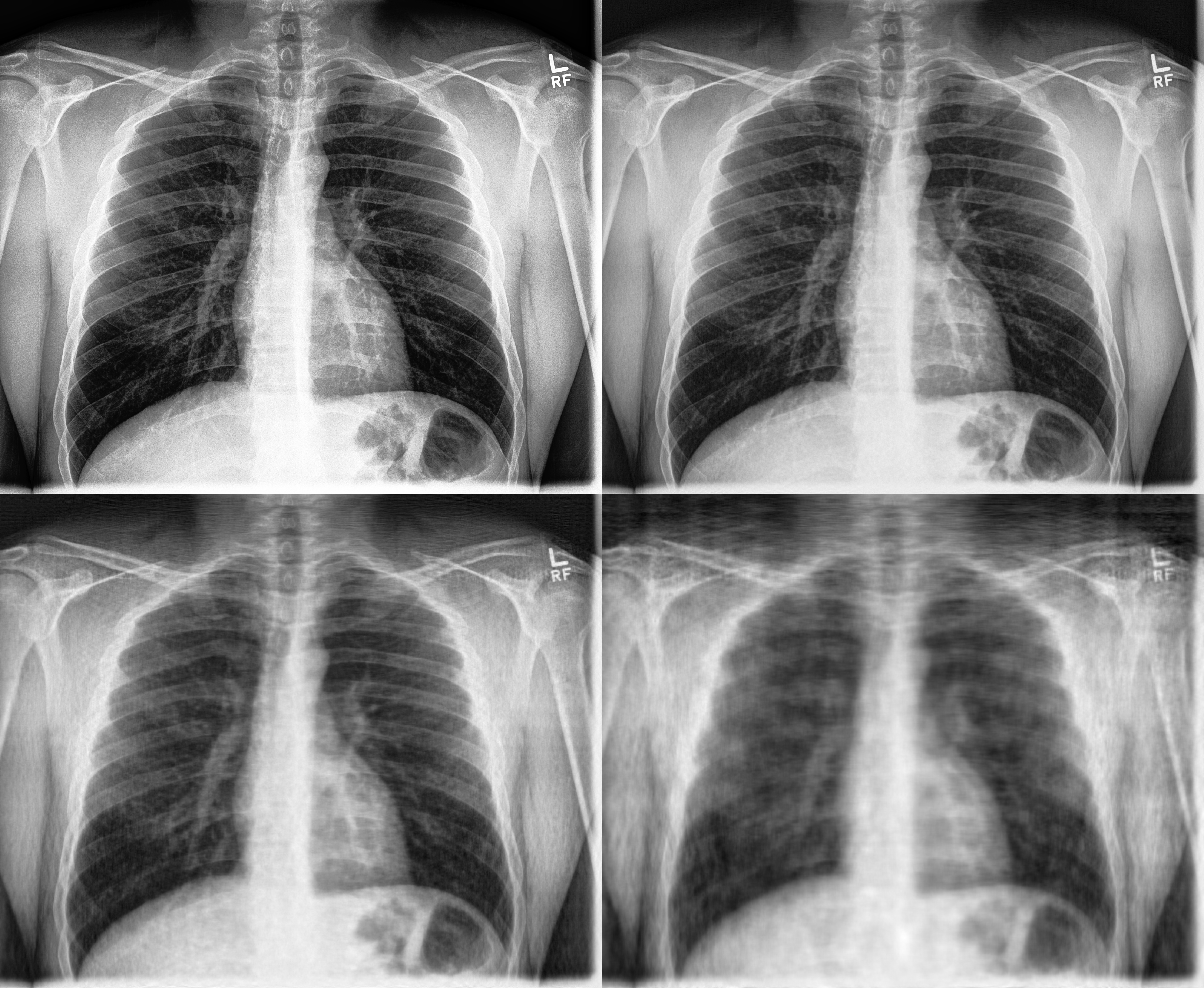}
\caption{Example reconstructions of a radiograph from the CheXpert dataset. Reduced dose radiographs were approximated by reducing the number used in the reconstruction. At the top right, the 1024-component reconstruction is virtually indistinguishable from the original (top left). The 512- and 256-component reconstructions at the bottom left and right, respectively, exhibit obviously reduced quality.}
\label{fig:recon}
\end{figure*}

Ensembles of lower-dose, mid-dose, and higher-dose radiographs were simulated by reconstructing CheXpert images using only 1024, 512, or 256 principle components in Equation~\ref{eq:recon}. A example is shown in Figure~\ref{fig:recon}. Each version of ``no finding'' archetypes and ``pneumonia indicated'' subjects were encoded into a deep feature space, and the KL-divergence computed as described in the methods. The result is shown in Table~\ref{tab:chexpert_kld}. The sign of the effect is reasonable; as reconstruction quality decreases, the divergence from the original image comprising all components increases. The scaling is non-linear, which makes sense in that there is no reason to assume that half the components means half the visual quality, however, it also is not clear how to use the divergence values to create such a scale. For example, the divergence between mid-quality no-finding and best-quality pneumonia is about the same as that between mid-quality no-finding and best-quality no-finding. In other words, the divergence is unable to distinguish typical reconstruction error (intra-ensemble variance) from inter-ensemble variance between the classes. Also, note that the divergence decreases down the diagonal. This makes sense because subtle features of the original image are adulterated in the poorer reconstructions. But that means that lowering quality has the same effect as increasing distribution overlap. Thus, for example, if a computed super-resolution network were to have mapped the 256-component images to 1024-component images, it is not clear how much decrease in divergence should be demanded or what the clinical meaning---e.g., intra- vs inter-class difference---of small differences in divergences is.

\begin{table}[ht]
\begin{center}
\caption{Kullback-Leibler divergence between the subject (S, row) and archetype (A, column) deep-feature point distributions for various reconstructions of CheXpert images.}
\label{tab:chexpert_kld}
\begin{tabular}{|c|c c c c|c|}
\hline
    \diagbox{subj}{arch} & all PCs & 1024 & 512 & 256 & intra\\ \hline
    all PCs & 185.3 & 232.9 & 316.7 & 842.5 & 0\\
    1024 & 236.5 & 184.0 & 250.5 & 772.9 & 86.15\\
    512 & 350.3 & 284.0 & 181.2 & 380.6 & 244.0\\
    256 & 823.5  & 790.8 & 423.7 & 175.4 & 726.0\\ \hline
    intra & 0  & 131.8 & 355.5 & 933.1 & NA \\ \hline
\end{tabular}
\end{center}
\end{table}

\begin{figure}[bth]
\begin{center}
\includegraphics[width=0.45\textwidth]{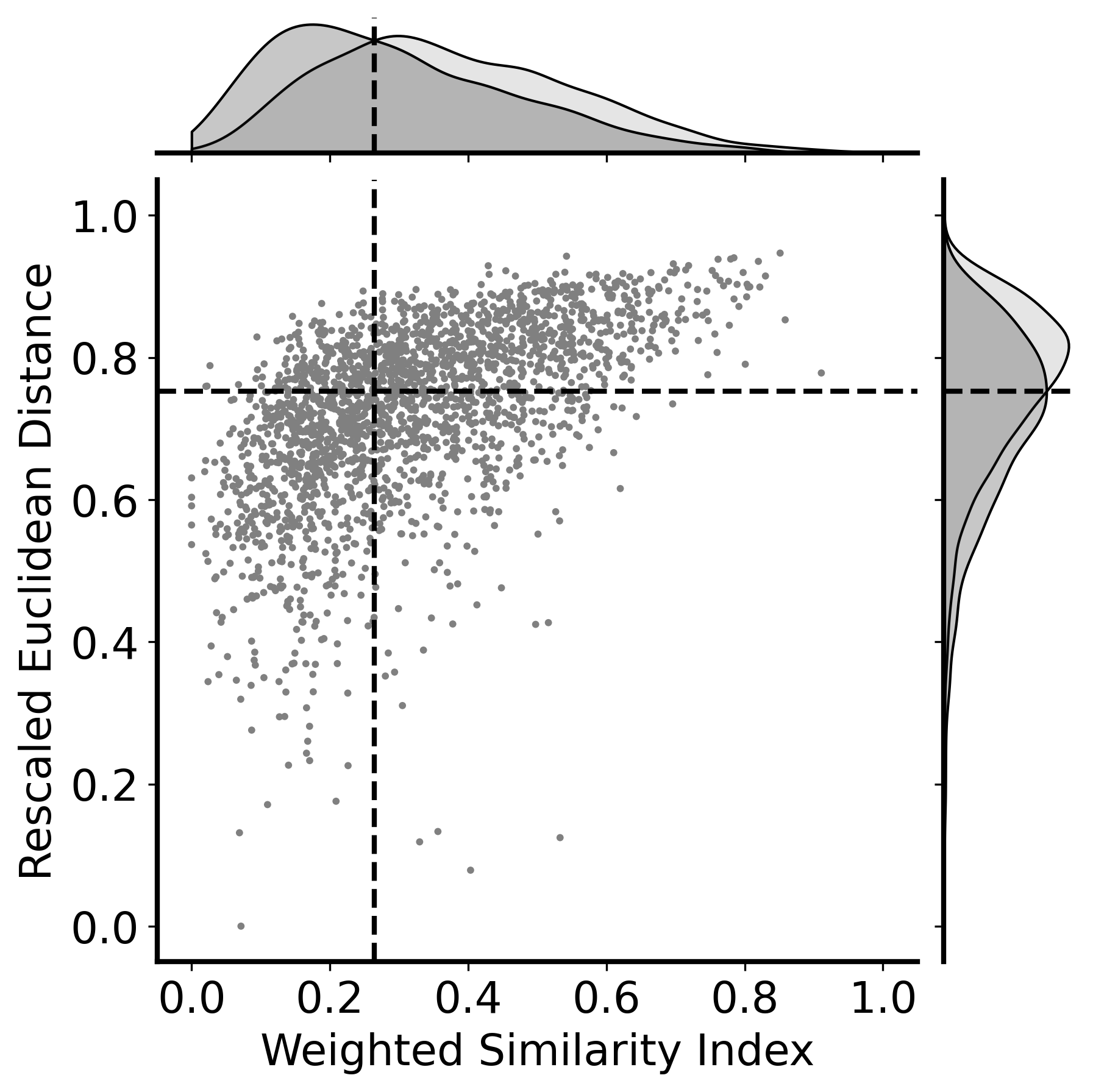}
\caption{Scatterplot of the distance-based measure vs the tally-based measure for a large number of inter-class pairs (no finding archetypes, pneumonia subjects). The plots at the top an right are kernel density estimates of the \emph{intra}-class similarity values (pneumonia-pneumonia is in darker gray). The dashed lines indicate the intersection of the univariate distributions.}
\label{fig:dist_vs_sim}
\end{center}
\end{figure}

Pairs of 1024-component pneumonia positive subjects were compared to no finding archetypes as follows. The same 60 conventional features employed throughout were computed for each image. As in an earlier example, a random forest classifier between subjects and archetypes was computed ($N$=2763, observed accuracy 70\%), and the component of the feature-weighting vector corresponding to the least important variable set to 1 and all other components set proportionally to variable importance. The bounding tolerance vectors ($\vec{l}$ and $\vec{u}$) were computed as follows. For each feature, the kernel density estimate for archetypes and subjects were plotted on the same axes such that the intersections can be found numerically. The feature tolerance was set to the interval where the probability of belonging to the archetypes is greatest. That is, if only that one feature were used to support the decision, the interval is defined to be where the probability of belonging to the archetype is greater. The WSI and Euclidean distance were computed for numerous random image pairings where at least 40 features appear in the universal feature set. The results are shown in Figure~\ref{fig:dist_vs_sim}. Note that the distance has been normalized and subtracted from one such that each measure has the same meaning (e.g., greater value means more similar). The overall width of the WSI distribution is greater, but that is largely because the distance measure yields outliers. Regardless of the absolute value, disagreement between the competing measures can be seen in at least two ways. First, note that the highest density region of WSI values for random pneumonia-pneumonia pairs (top, darker gray) is centered near 0.18, which indicates that pairs  most frequently are \emph{dis}similar. However, the distance-based measure shows that those same pairs most frequently pairs are similar ($\approx$0.75). Second, note the upper-left quadrant as defined by the dashed lines. Here, the tally-based measure indicates more dissimilar than similar (lower values), while the distance-based measure indicates more similar than dissimilar (higher values). The converse is seen the lower-right quadrant. The upshot is that for reasonable measure rescalings and, regardless of how the decision boundaries are defined, one can expect the distance-based measure and the tally-based to frequently disagree.

\subsubsection{Comparing Reconstructions}

\begin{figure}[bth]
\begin{center}
\includegraphics[width=0.475\textwidth]{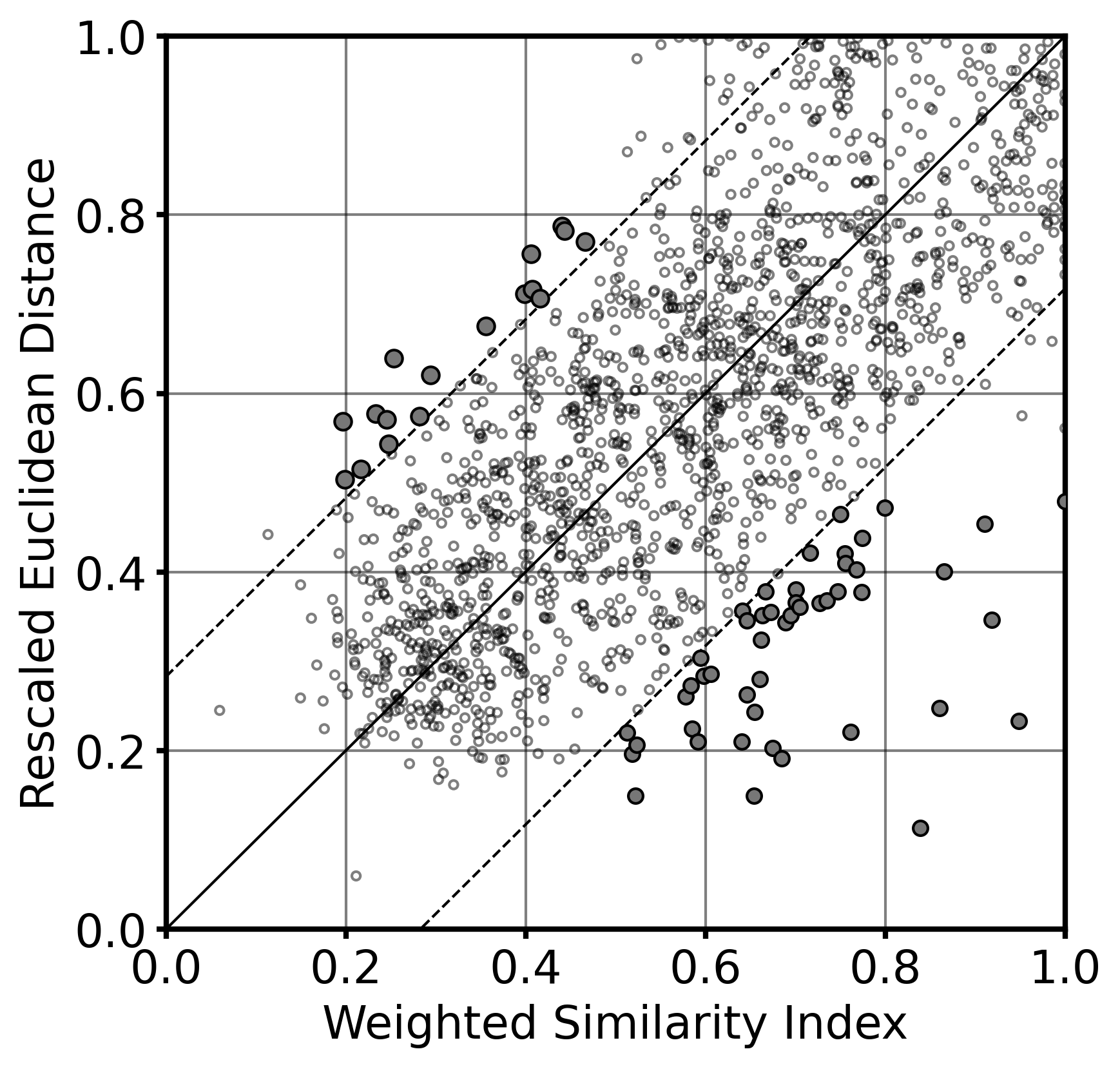}
\caption{There is clear disagreement between distance-based and tally-based measures of similarity when comparing paired versions of image reconstruction of CheXpert images; agreement is closeness to the line of identity. Many comparisons are outside a fairly large, stipulated measurement error ($\pm$0.2, dashed lines). Filled dots indicate extreme disagreement where the competing measures indicate the opposite conclusion about the similarity of two image reconstructions.}
\label{agreement_plot}
\end{center}
\end{figure}

To further investigate the potential for disagreement among similarity measures, a scenario where stark differences between comparands are expected was considered. Ensembles of lower-dose radiographs and higher-dose radiographs were simulated by reconstructing $N$=1646 ``no finding'' CheXpert images using only 256 or 1024 principle components in Equation~\ref{eq:recon}. The images are a holdout set from those used to computed the principal components (see Methods). The reconstructions are paired such that the 256-component reconstruction is an obviously lesser-quality version of the 1024-component reconstruction (see Figure~\ref{fig:recon}). The same 60 features used before were computed for each version. Because the difference in reconstructions is visually stark, unambiguous differences in feature space are expected. The Euclidean distance in feature space between each 256-1024 point pair was computed. The WSI was also computed using the same tolerances derived in the previous example. For ready comparison, the Euclidean distance was non-linearly rescaled to the same sign, range, and distribution of the WSI. That is, 0 means ``most dissimilar'' and 1 means ``most similar'' in both measures. The rescaled distance is plotted against the WSI in Figure~\ref{agreement_plot}. The dashed lines in Figure~\ref{agreement_plot} indicate plus or minus 0.2 from the line of identity. These boundaries effectively correspond to error bars around empirically measured similarity values. The large solid dots thus indicate comparisons where the competing measurements indicate the opposite conclusion. For example, one indicates ``more similar than dissimilar'' (value $>$ 0.5) while the other indicates ``more dissimilar than similar,'' (value $<$ 0.5). At 0.2 measurement error, only 4\% of the comparisons unambiguously disagree; however, if one has greater confidence in their measurements, such that a difference of 0.1 is considered distinguishable from measurement error, then 14\% of the comparisons unambiguously disagree. In other words, as ones confidence in the resolution of the distance-based measure increases, the agreement with the tally-based measure can decrease. Thus, with non-negligible frequency, when the intuitive and interpretable tally-based method indicates that two versions of image reconstruction are very similar, the distance-based measure indicates those same versions are dissimilar. This is an example the distance having not only an ambiguous effect size, but also ambiguous effect sign.


\section{Discussion}

\subsection{Some Practical Caveats}

There is an enormous literature on the taxonomy and comparison of so-called ``multi-dimensional entities'' that exhibit a large set of sophisticated, and often difficult to describe, features (see Refs. \cite{stolte2024methods} and \cite{ma2021image} for reviews). The upshot is that unambiguous criteria describing exactly how two things are to be declared (dis)similar must be stipulated prior to comparison; these criteria, together, are what Wittgenstein called a ``projection rule'' \cite{wittgenstein1930}. Experience-informed feature choice is perfectly consistent with scientific and engineering principles. If one were tasked with declaring that a complicated electro-mechanical device such as a pacemaker ready is for use, one would not employ just any arbitrary features but, instead, would have design specifications for each device component. Furthermore, one would not attempt to sum the manufacturing tolerances of all components into a single number (i.e., a distance) but, instead, would demand that each component be within its particular pre-specified tolerance or be rejected. This is precisely what we propose via our implementation of Tversky's method.

Because the weighted similarity index is a tally, it is a discrete random variable. This is not necessarily a detriment, but one should be aware when analyzing or comparing index values. One should also be mindful of the effect of missing features when comparing specific pairs drawn from ensembles. For example, suppose a feature set relevant to chest radiography includes a ``has pacemaker'' feature, $f_1$. It is possible that neither of two randomly drawn comparands exhibits $f_1$; one must explicate how this contingency is to be handled. It might be completely reasonable for the particular analysis to say that the comparands agree in that neither exhibits $f_1$; however, one could abuse this scheme by including a raft of rare or superfluous features to inflate the similarity between comparands. It also could be reasonable to ignore any feature that does not appear within either comparand at hand, but this could be misleading as well. For example, suppose one employs 1024 deep features, but only a very small subset of those happen to appear within tolerance for two particular radiographs. The tally of shared features might be 100\% because the comparands only have to agree in relatively few ways. Again, this is not necessarily a detriment, but one should be aware that the most unusual cases in an arbitrary feature space also might be the cases with the most extreme similarity values.

Unlike distances or divergences, the weighted similarity index is bounded between 0 and 1, and naturally scaled such that twice as great implies twice as similar. While these properties enable ready comparison of index values, one must still consider which features were employed before taking any value too seriously. For example, two radiographs may be 95\% similar via grayscale image moments, but only 80\% similar via radiologist-observed features. It would be up to the user to decide which value should be considered the better measure. When comparands are declared similar via \emph{all relevant} feature families, one might reasonably declare the comparands to be ``generally similar.'' Thus, Tversky's method is generalized by the discovery of relevant feature manifolds, and not necessarily by the expansion of feature space to greater range or dimension.

\subsection{Self-similarity Might Indicate One Form of Hallucination}

In Figure~\ref{boxplot_intra-wsi}, it is seen that some generative networks made images that tested as more dissimilar than would be expected from the variance of the training data. One way this could happen is if the network made the same very poor images repeatedly, however, this is assumed unlikely for a well-monitored, well-converged modern network. Another way is if the network made new classes of images not seen in the training data; this could be the case for an unconditional model that effectively interpolates new hybrid images that are not exactly any of the classes given in the data \cite{deshpande2024method}. Another explanation is that the data were augmented in a way inconsistent with the original data. In the AAPM challenge, every training image is of a horizontally asymmetric shape with identical left-right orientation, despite substantial variance in overall size and appearance. If left-right reversed images were added in an attempt to augment the data, hybrids of the correct orientation and the reversed orientation are possible; these would be new classes not seen in the original training data. Thus, comparison of the self-similarity of the generated ensemble to the self-similarity of the original training data might be an indicator of whether inter-class interpolation or extra-data augmentation has occurred.

\section{Conclusion}

The evaluation of images wholly or partially synthesized by a generative model is expected to be an assessment of image similarity, and not the computation of a pixel-wise difference from ground truth. Tversky's well-established, set theoretic method of multi-dimensional comparison can be applied to biomedical images as a robust measure of image similarity, even when the features exhibited are purely visual, highly variable in appearance, and not described formulaically. When machine automation is important, the numerical features employed to define similarity need only be axiomatically relevant to the data and analysis task at hand, and not agnostic or those most generally applicable. Once numerical features are stipulated, a tally of those features within pre-specified tolerances gives robust and intuitive results, especially in comparison to traditional distance- or divergence-based measures of image similarity.

\section{Future Work}

One obvious extension of our work is to develop a rigorous scheme for choosing the features when it is not clear which are relevant to the data and task. We have successfully employed a feature set empirically derived from a statistical analysis of expansion coefficients (see Equation~\ref{eq:recon}), however, many more data sets and tasks must be studied in order to comment further about the effectiveness of that feature-derivation scheme.
\bibliographystyle{ieeetr}
\bibliography{references.bib}

\end{document}